\documentclass[aps,pra,eqsecnum,twocolumn]{revtex4}
\usepackage{graphicx}
\usepackage{bm}
\usepackage{amsmath,amssymb,amsthm,dsfont,bm,enumitem}
\usepackage{color}
\usepackage{wasysym}
\usepackage{ulem}
\usepackage[dvipsnames]{xcolor}

\usepackage[applemac]{inputenc}

\usepackage[colorlinks=true,urlcolor=blue,citecolor=blue,linkcolor=blue]{hyperref}

\begin{document}
\preprint{}
\title{Single-measurement Bell analysis}
\author{Alfredo Luis}
\email{alluis@fis.ucm.es}
\homepage{https://sites.google.com/ucm.es/alfredo/inicio}
\affiliation{Departamento de \'{O}ptica, Facultad de Ciencias
F\'{\i}sicas, Universidad Complutense, 28040 Madrid, Spain}
\date{\today}

\begin{abstract}
We examine the satisfaction of Bell criteria for single realizations of quantum systems. This is possible via the joint noisy measurement of all observables involved in the Bell test.We readily find that every outcome violates Bell bounds for local hidden variables models. This agrees with the idea that to reveal nonclassical effects a necessary condition is that the measuring scheme itself must be nonclassical.
\end{abstract}
\maketitle

\section{Introduction}

 Local hidden-variable theories and Bell-like tests provide a powerful tool for investigating fundamental concepts at the very heart of quantum physics \cite{LB90,JB64,WW,AF82,AR15,BKO16,CH74,CHSH69,MA84,MC}. To this end, a key ingredient is the measurement of incompatible observables. This requires to combine results from different experimental realizations, that involve different physical contexts and that must be measured at different times. On the other hand, the derivation of the Bell-type bounds demands a unique probability distribution of hidden variables to cover all experimental realizations at all times. This assumption is not straightforward, since one may consider more physically motivated that hidden variables would depend on the observables being measured, given that the experimental settings are different \cite{MA84,MC,AK00,HP04,AM08,TN11,AK14,JCh17,NV62}. But in such a case, no meaningful bound might be derived.

This serious difficulty can be avoided by addressing practical schemes where all the observables involved in the Bell test are to be determined from a single experimental arrangement, where all measurements required are performed at once. This can be done via a noisy joint measurement \cite{MAL20,VRA23}. The key point is to design the  joint measurement such that its statistics provides enough information about all the observables involved in the corresponding Bell test, so that their exact statistics can be obtained via a simple data inversion from the measured data \cite{MAL22,MM89,WMM02,WMM14,PB87, AL16,LM17,GBAL18}. In particular, in Ref. \cite{MAL20} we have shown that it is possible to retrieve a joint distribution for all observables that fails to be a probability distribution by taking negative values exactly when the system state fails to satisfy the Bell theorem, in full agreement with Fine's theorem \cite{AF82}. All this within a unique statistical context as required in the Bell theorem.

Besides, the joint measurement provides a further interesting possibility. This is the test of Bell-type bounds for single realizations of the system. This is because every single realization has enough information about all observables involved in the test as to examine the satisfaction of the Bell bounds. In this work we examine this possibility in terms of the scheme in Ref. \cite{MAL20}. A related project has already been carried out experimentally in Ref. \cite{VRA23} under a different scheme that in principle does not address the retrieval of the exact statistics.We can also mention the approach in Ref. \cite{AHQ20} showing the possibility of single-shot rejection of local hidden variables theories. Note that in general single outcomes do not provide enough information to allow a complete tomographic reconstruction of the state, which  generally requires repeated realizations and the measurement of other observables not involved in the Bell test, and which would go against the main objective of this work. 

It is worth noting that our approach, focusing on probability and statistics, is completely independent of the theory, classical or quantum, within which such probabilities are derived. No quantum mechanical rules are involved. Precisely because of this, this method can reveal non-classical properties in the form of observed statistics that do not comply with the  constraints imposed by hidden variable models. 

So we propose an entirely new application of a previous scheme that has already proven its usefulness in Bell-type analysis. Given the relevance of the subject regarding fundamental concepts of quantum theory we think that the proposal in this work is novel enough and deserves investigation. 

\section{Method}

\subsection{Local hidden variable model}

As usual, our system will be made of two not causally connected subsystems $A$ and $B$. We consider Bell tests in terms of pairs of observables defined in each subsystem. These are  $X,Y$ in the subsystem $A$, and $U,V$ in the subsystem $B$, al them being dichotomic, taking values $x,y, u,v = \pm 1$. The local hidden variable model to be tested by the Bell-type criteria prescribes the following form for the joint statistics of a pair of $A$ and $B$ observables 
\begin{equation}
\label{LHVM}
    p_{Z,W} (z,w) = \int d\lambda p_Z (z|\lambda )  p_W (w|\lambda ) P_\Lambda (\lambda) ,
\end{equation}
for observables $Z=X,Y$, $W=U,V$ corresponding to variables $z=x,y$, and $w = u,v$, being $\lambda$ the hidden variables with probability distribution $P_\Lambda (\lambda)$, and  $p_Z (z|\lambda )$,  $p_W (w|\lambda )$ are the corresponding conditional probabilities. 

As a simple convenient Bell-type test we may consider the Clauser-Horne-Shimony-Holt (CHSH) inequality \cite{CHSH69} in terms of correlations of pairs of $A$ and $B$ observables. More specifically, we consider the magnitude $S$ defined as
\begin{equation}
\label{Save}
S = \langle xu \rangle - \langle xv \rangle  + \langle yu \rangle + \langle yv \rangle ,
\end{equation}
where 
\begin{equation}
\langle zw \rangle = \sum_{z,w} z w  p_{Z,W} (z,w) ,
\end{equation}
so that in a local hidden-variable theory of the form (\ref{LHVM}) it holds that 
\begin{equation}
\label{ccm}
    |S| \leq 2 .
\end{equation}

As a further Bell-type test we may consider the Fine criterion in the form of a combination of probabilities for outcomes of all the $A$, $B$ observables \cite{CH74,AF82}
\begin{eqnarray}
\label{BCH}
&C (\xi) = p_{X,U} (x,u) - p_{X,V} (x,v) +\\
&p_{Y,U} (y,u) + p_{Y,V} (y,v)- p_Y (y) - p_U (u) ,\nonumber
\end{eqnarray}
where  $\xi$ denotes the set of the four variables $\xi = ( x,y,u,v)$. For the above classical-like local hidden-variable model (\ref{LHVM}) it holds that
\begin{equation}
\label{CC}
        0 \geq C \geq -1 .
\end{equation}

As mentioned in the introduction, a key point is that the measurement of incompatible observables, such as $X$, $Y$ and $U$, $V$, require different experimental realizations, and so different physical contexts. On the other hand, the derivation of the Bell-type bounds demand a unique probability distribution of hidden variables, say $P_\Lambda (\lambda)$, to cover all experimental realizations at once. The bounds shown above would not exist if we would consider a different $\lambda$'s and different $P_\Lambda (\lambda |Z,W)$ depending on the observables $Z$, $W$ being measured, which would seem more physically motivated given that the experimental settings are different. 

This serious difficulty can be avoided by addressing practical schemes where all the observables involved in the Bell test are to be determined from a single experimental arrangement where all measurements required are performed at once. This can be done via a noisy joint measurement \cite{MAL20,VRA23} .

\subsection{Noisy simultaneous joint measurement and inversion}

We consider a noisy simultaneous joint measurement of the observables $X,Y,U,V$, leading to a jointly observed statistics $\tilde{p} ( \xi^\prime |\rho )$ where $\xi^\prime$ denotes the set of four measurement outcomes $\xi^\prime = (x^\prime,y^\prime,u^\prime,v^\prime)$, and $\rho$ represents the actual system state at work, either in the classical or in the quantum theories, it makes no difference for the method. 

Let us preset here the general method of data inversion to obtain noiseless results from the noisy data. Particular expressions will be given later in the typical situation where all variables are dichotomic, as it corresponds to qubit system realizations and the customary assumptions of the Bell test presented above. 

\bigskip

The key point is that marginal statistics $\tilde{p}_K (\kappa^\prime)$, where $\kappa^\prime$ may be  any of the measurement outcomes $x^\prime,y^\prime,u^\prime,v^\prime$, say $x^\prime$ for instance,
\begin{equation}
\label{marg}
\tilde{p}_X (x^\prime) =   \sum_{y^\prime,u^\prime,v^\prime} \tilde{p} ( x^\prime |\rho ) ,
\end{equation}
are assumed to provide complete information about the corresponding observable $K =X,Y,U,V$, so that the exact statistics $p_K (\kappa)$ of $K$ can be obtained from $\tilde{p}_K (\kappa^\prime)$. This is that for each $K$ there are state-independent functions $p_K (\kappa | \kappa^\prime) $ so that the exact  $p_K (\kappa)$ is obtained from the marginal $\tilde{p}_K (\kappa^\prime)$ as 
\begin{equation}
\label{pW}
p_K (\kappa) =   \sum_{\kappa^\prime } p_K (\kappa | \kappa^\prime) \tilde{p}_K (\kappa^\prime) .
\end{equation}

\bigskip

We can collect all these inversions in a single joint distribution $p (\xi |\rho )$, where $\xi = ( x,y,u,v)$, defined as
\begin{equation}
\label{chr}
    p (\xi |\rho ) = \sum_{\xi^\prime} p (\xi |\xi^\prime ) \tilde{p} ( \xi^\prime |\rho ) ,
\end{equation}
where
\begin{equation}
\label{pchichiprime}
    p (\xi |\xi^\prime )  = p_X (x | x^\prime )p_Y (y | y^\prime )p_U (u | u^\prime )p_V (v | v^\prime ) .
\end{equation}

Regarding the statistical nature of these quantities, we must distinguish $p (\xi |\rho )$ and $\tilde{p} (\xi^\prime |\rho)$ from $p (\xi |\xi^\prime )$. This last one $p (\xi |\xi^\prime )$, which is a relation between statistics (\ref{chr}), is a state-independent deterministic nonrandom function fully determined once for all once we know the measurement to be performed. On the other hand, $p (\xi |\rho )$ and $\tilde{p} (\xi^\prime |\rho)$ are of random statistical nature depending on the state $\rho$ considered. To be determined one would need to perform a large enough number of single-shot measurements to collect the complete statistics.

We have to stress that in classical physics this inversion procedure always leads to a proper joint probability distribution, this is $p(\xi |\rho) \geq 0$, since in classical physics such a proper probability distribution exists for all combination of observables \cite{AL16,LM17}. However in the quantum case this is no longer granted when the are incompatible observables, so the lack of positivity of $p(\xi |\rho)$ is a signature of nonclassical behavior, in agreement with the Fine theorem in Ref. \cite{AF82}. We recall that $\tilde{p} (\xi^\prime |\rho)$ is always a non negative probability distribution by virtue of being the statistics of the actual measurement performed. Consequently, after Eq. (\ref{chr}), the lack of positivity of $p(\xi |\rho)$ requires a lack of positivity of $ p (\xi |\xi^\prime )$.  This naturally can be related to the nonclassical features displayed by the Bell-tests measurements, as shown in Ref. \cite{MAL20}.

\subsection{Single shots}

We may regard $ p (\xi |\xi^\prime )$ as the inferred conditional joint distribution for the Bell-test variables $\xi$  associated to the single outcome $\xi^\prime$ of the joint measurement. This is the analog of $p (\xi |\rho)$ but associated, not with the system state $\rho$, but with the outcome $\xi^\prime$. This is actually the core of the novel single-shot aspect addressed in this work. This is that $p (\xi |\xi^\prime )$ is as much as we can say of the physics involved in the single result $\xi^\prime$. 

Let us apply this idea to the CHSH test in Eq. (\ref{Save}), that can be formulated as the mean value of the following combination of system variables 
\begin{equation}
s(\xi) = xu - xv + yu + yv .
\end{equation}
So, according to the above reasoning, the value of the CHSH test when the single-shot outcome is $\xi^\prime$, which we shall call $S(\xi^\prime)$, is given by
\begin{equation}
\label{Sxi}
S(\xi^\prime) = \sum_\xi  s(\xi) p(\xi |\xi^\prime ) .
\end{equation}
This interpretation of $p (\xi |\xi^\prime )$ is further supported by showing the the average of $S(\xi^\prime)$ under the measurement statistics $\tilde{p} (\xi^\prime |\rho )$ is precisely the same $S$ in Eq. (\ref{Save}), this is after Eq. (\ref{chr}) 
\begin{equation}
\label{fchsh}
S = \sum_{\xi^\prime}  S(\xi^\prime) \tilde{p} (\xi^\prime |\rho ) = \sum_\xi s(\xi) p(\xi |\rho ),
\end{equation}
that is the standard value of the CHSH test.

We apply a similar reasoning regarding the single-shot version of the Fine criterion in Eq. (\ref{BCH}), which we shall call $C(\xi |\xi^\prime)$, which is given by the very same expression (\ref{BCH}), but where the $p_{Z,W} (z,w)$ are replaced by their counterparts in terms of the conditional inferred joint distribution $ p(\xi |\xi^\prime )$, this is for example
\begin{equation}
    p_{X,U} (x,u) \rightarrow  p_{X,U} (x,u|\xi^\prime)= \sum_{y,v=\pm 1} p(\xi |\xi^\prime ).
\end{equation}
Here again, the  exact $C(\xi)$ in Eq. (\ref{BCH}) is retrieved as an average over the $\xi^\prime$ statistics $\tilde{p} (\xi^\prime |\rho )$ so that
\begin{equation}
  C (\xi) = \sum_{\xi^\prime} C(\xi | \xi^\prime) \tilde{p} (\xi^\prime |\rho ) ,
\end{equation}
which supports our interpretation of $C(\xi | \xi^\prime)$ as a single-shot realization of the test. 

\subsection{Dichotomic case}

In order to proceed further we may consider the usual two-dimensional scenario for subsystems $A$ and $B$  where both set of system and measured variables are dichotomic, this is $x,y,u,v =\pm 1$ as well as $x^\prime,y^\prime,u^\prime,v^\prime =\pm 1$. This may be the case for example of the polarization of two photons in two distinguishable field modes.

In this case, given the small dimension of the system space, there is a simple relation between the noisy marginals $\tilde{p}_K (\kappa^\prime)$ and the noiseless ones $p_K (\kappa)$, for $K=X,Y,U,V$. Taking into account a natural unbiasedness requirement regarding that if  $p_K (\kappa)$ is uniform then so is $\tilde{p}_K (\kappa^\prime)$ \cite{YLLO10}, and that $\kappa,\kappa^\prime = \pm 1$, this relation must be of the form
\begin{equation}
\label{nn}
\tilde{p}_K (\kappa^\prime) = \frac{1}{2} \sum_\kappa \left  ( 1 +  \gamma_K \kappa \kappa^\prime  \right ) p_K (\kappa) , 
\end{equation}
where $\gamma_K$ are real factors with $|\gamma_K| \leq 1$. These factors are expressing the accuracy in the observation so that the noisy mean value of each observable $K$, say $\langle \tilde{K} \rangle$, 
\begin{equation}
    \langle \tilde{K} \rangle = \sum_{\kappa^\prime} \kappa^\prime \tilde{p}_K (\kappa^\prime | \rho)  ,
\end{equation}
is related to the exact one
\begin{equation}
    \langle K \rangle = \sum_{\kappa} \kappa  p_K (\kappa | \rho),
\end{equation}
in the form 
\begin{equation}
\langle \tilde{K} \rangle = \gamma_K \langle K \rangle .
\end{equation}
Note that for dichotomic variables the variance is determined by the mean value 
\begin{equation}
\Delta^2 K = 1 - \langle K \rangle^2, 
\end{equation}
so $\Delta^2 \tilde{K}$ is always larger than $\Delta^2 K$, where the lesser the $|\gamma_K |$, the larger is the difference. 

The factors $\gamma_K$ are state-independent being determined just by the particular measurement arrangement considered. They are not free parameters since they are constrained so that the joint $ \tilde{p} (\xi^\prime | \rho) $ satisfies all necessary requirements to be a {\it bona fide} probability distribution, including positive semidefiniteness.  In some simple but meaningful realizations this means that \cite{MAL22}
\begin{equation}
    \gamma_X^2+ \gamma_Y^2 \leq 1, \qquad \gamma_U^2+ \gamma_V^2 \leq 1 .
\end{equation}
In this case, the functions $p_K (\kappa | \kappa^\prime) $ in Eq. (\ref{pW}) that perform the inversion of Eq. (\ref{nn}) are 
\begin{equation}
\label{pp}
p_K (\kappa | \kappa^\prime) = \frac{1}{2} \left ( 1 + \frac{\kappa \kappa^\prime}{ \gamma_K}  \right ) .
\end{equation}

It is worth noting that, besides obtaining the correct one-observable marginals by construction, $p(\xi|\rho)$ also contains the exact two-observable statistics for any pair of one observable in $A$ and another in $B$ \cite{MAL20}. This is for example 
\begin{equation}
\label{toep}
    p_{X,U} (x,u) =\sum_{y,v} p (\xi|\rho) .
\end{equation}

Let us note that relations in Eqs. (\ref{nn}) are central to this approach. This is because they are connecting original and noisy observables, and their inverse (\ref{pp}) provide the single-shot inverted distributions given the data (\ref{pchichiprime}). In principle, these relations work equally well in the classical and quantum theories, and can be verified experimentally. This is simply a matter of comparing the exact and noisy statistics of the corresponding observable, a procedure that in principle is independent of the theory.

This being the situation, a word of caution may be appropriate here, which may be always advantageous in the uncertain terrain between so dissimilar theories. The analysis presented here is a first approach to the issue, which certainly can and should be further investigated in relation to possible problems of interpretation and loopholes.

\section{Single-shot CHSH test}

For the simple dichotomic case we can easily compute the single-shot Bell criterion $S(\xi^\prime)$ value in Eq. (\ref{Sxi}) after Eqs. (\ref{pchichiprime}) and (\ref{pp}), leading to 
\begin{equation}
    S(\xi^\prime ) = \frac{\gamma_Y \gamma_V x^\prime u^\prime - \gamma_Y \gamma_U x^\prime v^\prime + \gamma_X \gamma_V y^\prime u^\prime + \gamma_X \gamma_U y^\prime v^\prime}{\gamma_X \gamma_Y \gamma_U \gamma_V} .
\end{equation}
We stress that outcomes are not redefined at any instance, as throughout they are always +1 and -1, both for $(x, y, u, v)$ and $(x^\prime, y^\prime, u^\prime, v^\prime)$. The focus of this process of data inversion for noise removal is on statistics, not on outcomes. Nevertheless, for dichotomic variables full statistics are just determined by their mean values so the final expressions must be necessarily simple.

To simplify the analysis let us consider the natural case where all the gamma factors are equal, $\gamma_X = \gamma_Y = \gamma_U = \gamma_V = \gamma$
leading to 
\begin{equation}
    S(\xi^\prime ) = \frac{1}{\gamma^2} \left (x^\prime u^\prime - x^\prime v^\prime + y^\prime u^\prime + y^\prime v^\prime \right ) .
\end{equation}
Given that $x^\prime,y^\prime,u^\prime,v^\prime = \pm 1$ it can be easily seen that 
\begin{equation}
|S (\xi^\prime )| = \frac{2}{\gamma^2}  >2 \quad \forall \xi^\prime ,
\end{equation}
so the inequality (\ref{ccm}) of the hidden-variable model is clearly violated for every outcome $ \xi^\prime$. 

Let us stress that in this work we have developed the tools to apply the test in Eq. (\ref{ccm}) to single realizations by finding the single-system realizations $S(\xi^\prime)$ of $S$. We have just shown that the single-shot violations hold for every outcome $\xi^\prime$ and every $\rho$. This independence on $\rho$ seems natural since $\rho$ determines how the outcomes $\xi^\prime$ are distributed, but the physical meaning of each outcome $\xi^\prime$ must be solely determined by the measurement process itself, and this includes whether the arrangement is able to disclose violations of local hidden-variables models.

Naturally, the final ensemble result $S$ in Eq. (\ref{fchsh}), and whether or not it satisfies local hidden-variable bounds, depends on $\rho$ through the $\tilde{p} (\xi^\prime |\rho )$ probability distribution, so that $S$ may satisfy the hidden-variable model even if every single-shot violates it. Note that the ensemble averaged $S$ arises from the average in the first equality of Eq. (\ref{fchsh}) where the single-shot $S(\xi^\prime)$ contributes as quantities with sign, not just modulus, so that $S$ may not violate the modulus bound although all $S(\xi^\prime)$ do.

It is known that ensemble-averages violations of the modulus bound for $S$ are equivalent to negativities of $p (\xi|\rho)$ \cite{AF82,MAL20}, so in the same terms violations of the modulus bound by $S(\xi^\prime)$ are given by negativities of $p(\xi|\xi^\prime)$. This is closely related to the fact that the measuring scheme is able to disclose potential nonclassical properties of the system state, as we have already shown regarding the Bell tests in Ref. \cite{LA20}. This enforces the fruitful idea that quantum statistics is a combination of system and apparatus contributions, and it holds that to reveal nonclassical properties in the quantum realm a necessary condition is that the measuring scheme itself must be nonclassical, where this is actually the case for most detection arrangements \cite{LA20,RL09,LR11}.

\section{Single-shot Fine criterion}

After Eqs. (\ref{pchichiprime}) and (\ref{pp}) we have
\begin{equation}
C(\xi |\xi^\prime) = -\frac{1}{2} - \frac{x x^\prime v v^\prime}{4 \gamma_X \gamma_V} + \frac{y y^\prime v v^\prime}{4 \gamma_Y \gamma_V} + \frac{x x^\prime u u^\prime}{4 \gamma_X \gamma_U} + \frac{y y^\prime u u^\prime}{4 \gamma_Y \gamma_U} .
\end{equation}
For simplicity we may once again assume that all gamma factors are equal, $\gamma_X = \gamma_Y = \gamma_U = \gamma_V = \gamma$, so that $C(\xi |\xi^\prime)$ can actually take only two values
\begin{equation}
C(\xi |\xi^\prime) = -\frac{1}{2} \pm \frac{1}{2\gamma^2} ,
\end{equation}
and both of them violate one of the bounds in Eq. (\ref{CC}) for all $\xi$ and $\xi^\prime$ since 
\begin{equation}
 -\frac{1}{2} + \frac{1}{2\gamma^2} >0, \quad  -\frac{1}{2} - \frac{1}{2\gamma^2} <-1 .
\end{equation}

\section{Conclusions}

We have examined the satisfaction of Bell criteria for single realizations of quantum systems via the noisy joint measurement of all observables involved, that in spite of being noisy contain full information about all the observables and their correlations. We have found that every outcome violates Bell bounds for local hidden variables models. This can be understood from the quantum nature of the detection processes and the fact that to reveal nonclassical properties a necessary condition is that the measuring scheme itself must be nonclassical \cite{LA20,RL09,LR11}.

\section*{ACKNOWLEDGMENTS} 
A. L. thanks Dr. F. Piacentini and Dr. L. Ares for fruitful discussions.

\end{document}